\documentclass[12pt]{iopart}
\usepackage[colorlinks=true,linkcolor=blue,anchorcolor=blue,citecolor=blue]{hyperref}
\usepackage{graphicx}
\usepackage{xcolor}
\usepackage{bm}
\usepackage{scrextend}
\begin{document}

\title{Quasiparticle multiplets and 5$f$ electronic correlation in prototypical plutonium borides}
\author{Haiyan Lu and Li Huang}
\address{Science and Technology on Surface Physics and Chemistry Laboratory, P.O. Box 9-35, Jiangyou 621908, China}
\ead{\mailto{hyluphys@163.com}}
\vspace{10pt}

\begin{abstract}
In this paper, we investigate the electronic structures of plutonium borides (PuB$_x$, $x$=1, 2, 6, 12) to uncover the fascinating bonding behavior and orbital dependent correlations of 5$f$ valence electrons by using the density functional theory combined with single-site dynamical mean-field method. 
We not only reproduce the correlated topological insulator of PuB$_6$, but also predict the metallicity in PuB$_x$ ($x$=1, 2, 12). It is found that the band structure, density of states, hybridization functions all indicate partially itinerant 5$f$ states in PuB$_x$ ($x$=1, 2, 6, 12). Especially, quasiparticle multiplets induced noteworthy valence state fluctuations implying the mixed-valence behavior of plutonium borides. Moreover, the itinerant degree of freedom for 5$f$ electrons in PuB$_x$ ($x$=1, 2, 12) is tuned by hybridization strength between 5$f$ states and conduction bands, which is affected by atomic distance in plutonium borides. Lastly, 5$f$ electronic correlations encoded in the electron self-energy functions demonstrate moderate 5$f$ electronic correlations in PuB$_6$ and orbital selective 5$f$ electronic correlations in PuB$_x$ ($x$=1, 2, 12). Consequently, the understanding of electronic structure and related crystal structure stability shall shed light on exploring novel 5$f$ electrons states and ongoing experiment study.
\end{abstract}
\vspace{2pc}
\noindent{\it Keywords\/}: dynamical mean-field method, electronic correlation, quasiparticle multiplets, mixed-valence behavior

\submitto{\JPCM}
\maketitle

\section{Introduction\label{sec:introduction}}
Actinide compounds with partially filled 5$f$ electronic shell exhibit abundant fantastic behavior of fundamental physical interests including heavy-fermion behaviors, unconventional superconductivity, nontrivial topology, complex magnetism and mixed-valence states~\cite{RevModPhys.81.235,PhysRevLett.108.017001,PhysRevLett.91.176401,Sarrao2002,Curro2005,Daghero2012,PhysRevLett.111.176404,PhysRevB.99.035104,PhysRevB.97.201114}. These perplexing phenomena are primarily attributed to the intricate electronic structure concerning strongly correlated 5$f$ electrons and hybridization between 5$f$ electrons and conduction bands. It is well known that plutonium (Pu) occupies the position on the edge between the itinerant and localized 5$f$ states~\cite{RevModPhys.81.235} of light and heavy actinides, Pu metal reveals multiple allotropic crystalline phases~\cite{savrasov:2001} and tremendous variations in thermophysical properties. Note that 5$f$ electron is extremely sensitive to external temperature, pressure and chemical doping, so the active 5$f$ electrons are prone to take in chemical bonding and form plenty of plutonium-based compounds~\cite{Bauer2015Plutonium}. Particularly, plutonium tetraboride (PuB$_4$) and plutonium hexaboride (PuB$_6$) have been reported to be promising correlated topological insulators~\cite{PhysRevLett.111.176404,PhysRevB.99.035104,PhysRevB.97.201114,PhysRevB.102.085150}.

\begin{figure}[ht]
\centering
\includegraphics[width=0.6\columnwidth]{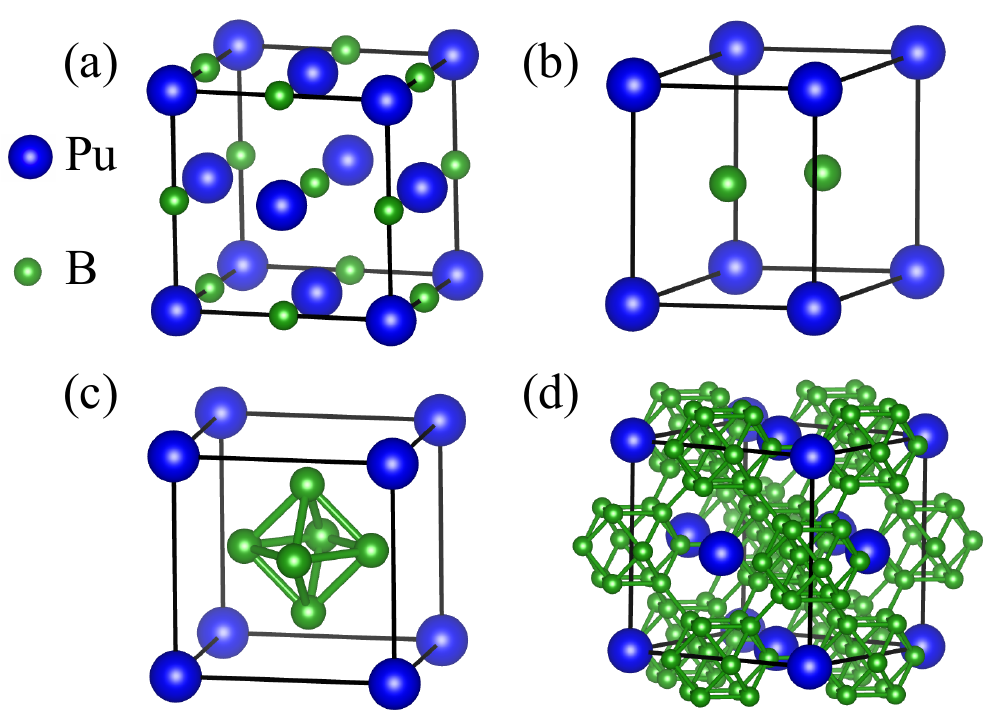}
\caption{(Color online). Crystal structure of plutonium borides, where blue and green spheres mean Pu and B atoms, respectively. (a) PuB. (b) PuB$_2$. (c) PuB$_6$. (d) PuB$_{12}$. Note that the atomic distances between Pu-Pu, Pu-B, B-B are the nearest neighbour.
\label{fig:struct}}
\end{figure}

\begin{table}[th]
\begin{center}
\caption{Crystal structure parameters of plutonium borides~\cite{McDonalda02824,Eick1965} PuB$_x$ ($x$=1, 2, 6, 12). \label{tab:param}}
\begin{tabular}{ccccc}
PuB$_x$ & Space group & $d_{\texttt{Pu-Pu}}$ ({\AA}) & $d_{\texttt{B-B}}$ ({\AA})& $d_{\texttt{Pu-B}}$ ({\AA}) \\
\hline 
PuB & $Fm$-3$m$ & 4.92 & 3.48 & 2.46 \\
PuB$_2$ & $P$6/$mmm$ & 3.19 & 1.84 & 2.70 \\
PuB$_6$ & $Pm$-3$m$ & 4.11 & 1.70 & 3.03 \\
PuB$_{12}$ & $Fm$-3$m$ & 5.29 & 1.76 & 2.79 \\
\hline
\end{tabular}
\end{center}
\end{table}

Plutonium borides are typical actinide borides (Th-, U-, Np-, Pu-, Am-borides~\cite{Eick1969}), which integrate the electronic, optical, mechanical and refractory properties of metal borides~\cite{borides2017}. In 1960s plutonium borides were sucessfully synthesized in spite of the intractable radioactivity and toxicity. The high melting point ($\sim$2300 K~\cite{Eick1965}) and refractory~\cite{Rogl1997} enable their potential use as nuclear fuels. Generally, plutonium borides stablize in PuB, PuB$_2$, PuB$_4$, PuB$_6$, PuB$_{12}$ and PuB$_{100}$ according to various boron-containings~\cite{McDonalda02824,Eick1965}. 
It is worth noting that PuB$_4$ crystalizes in the tetragonal ThB$_4$-type crystal structure with space group $P4/mbm$ (No. 127)~\cite{McDonalda02824,Bauer2015Plutonium,Rogl1997} and PuB$_{100}$ has a large number of boron atoms in the unit cell.
Here we only consider PuB$_x$ ($x$=1, 2, 6, 12) with relatively high crystal structure symmetry, whose crystallographic parameters measured by X-ray powder diffraction techniques are listed in table~\ref{tab:param}.
Plutonium monoboride (PuB)[Fig.~\ref{fig:struct}(a)] stablizes in a cubic NaCl-type structure (space group $Fm$-3$m$). It is the lowest boron-containing plutonium boride with the absence of uranium and thorium monoborides. Plutonium diboride[Fig.~\ref{fig:struct}(b)] takes a AlB$_2$-type hexagonal structure (space group $P$6/$mmm$) where layered metal atoms interposed in the two-dimensional hexagonal networks formed by boron atoms. Plutonium hexaboride[Fig.~\ref{fig:struct}(c)] has a cubic structure (space group $Pm$-3$m$), where Pu atoms establish a skeleton with boron octahedra locating at the cubic corners. Plutonium decoboride[Fig.~\ref{fig:struct}(d)] crystalizes in a cubic structure (space group $Fm$-3$m$), where Pu ions and B$_{12}$ units arrange as NaCl-type. Each Pu ion locates in the center of a B$_{24}$ cubo-octahedron which associates B$_{12}$-polyhedra cages formulating a three-dimensional skeleton. In this sense, the basic structural units in plutonium borides are depicted in terms of the B-B bonding configurations, such as B$_2$-$sp^2$, B$_6$-octahedral, and B$_{12}$-cubeoctahedral boron atom clusters. Rich physics involving complicated electronic structure and exotic topological feature are expected because of the fantastic bonding behavior between Pu and B atoms.

Appreciable experimental progresses in plutonium borides have been achieved on the crystal structure, magnetism, transport, thermodynamics and reactivity~\cite{Rogl1997,LARROVUE1986487,BOROVIKOVA1986287,Chipaux132,Smith7883}. Available Pu-B phase diagram on temperature and boron composition parameter plane clarifies the stable condition and locates the melting points of plutonium borides~\cite{Rogl1997}, even though some early results still dispute the existence of PuB~\cite{LARROVUE1986487} owing to the various preparation methods. Besides, entropies have been measured for PuB$_2$, PuB$_4$ and PuB$_6$ at room temperature~\cite{BOROVIKOVA1986287}. It is reported that magnetic susceptibility of PuB$_2$ slightly depends on temperature~\cite{Chipaux132}. Whereas magnetic susceptibility of PuB$_4$ and PuB$_6$ rarely changes against temperature~\cite{Smith7883}, indicating different nature from PuB, PuB$_2$ and PuB$_{12}$ metal. Recent experimental advances confirmed the correlated topological insulator of PuB$_4$ through small magnetic moment of Pu atom, insulating like electrical transport and $^{239}$Pu nuclear magnetic resonance. Among several Pu-based compounds which have been predicted to be correlated topological insulators, PuB$_4$ and PuB$_6$ are archetypical prototypes to explore the underlying mechanism of the topological feature and strongly correlated 5$f$ electronic states.

PuB$_6$ has been addressed to exhibit topological feature and mixed-valence behavior for 5$f$ electrons within the combination of dynamical mean-field approach and density functional theory~\cite{PhysRevLett.111.176404}. Furthermore, PuB$_6$ is proposed to be Racah material with intermediate valence singlet ground state by utilizing local density approximation plus an exact diagonaliztion method~\cite{Shick15429}. Subsequently, the surface states signatures reflected topological feature are predicted in scanning tunneling spectroscopy and quasiparticle interference based on a multiorbital tight-binding model~\cite{PhysRevB.90.201106}. Lately, You Lv \emph{et al.} theoretically found that PuB$_{12}$ is the most stable compound among $A$B$_{12}$ ($A$=Th, U, Np, Pu, Am) and a candidate superhard material owing to the outstanding mechanical and thermodynamic properties~\cite{LV2018128}. Motivated by these findings, it is worth while to survey the electronic structure tuned by boron composition to build a comprehensive picture of strongly correlated 5$f$ electronic states, itinerant-localized dual nature and bonding behavior, so as to pave the way for future angle-resolved photoemission spectroscopy and de Haas-van Alphen (dHvA) oscillation experiments.

In these calculations, theoretical investigations mainly concentrate on PuB$_4$ and PuB$_6$, while experimental advances achieve pioneering results on topological insulator in PuB$_4$. However, present research on PuB, PuB$_2$ and PuB$_{12}$ are insufficient. Moreover, the long-standing issue of itinerant-localized dual nature for 5$f$ electrons which is closely linked with crystal stability has rarely been touched. Above all, the accurate electronic structure involving the strong 5$f$ electronic correlation, large spin-orbit coupling, and intricate crystal field splitting have not been sufficiently taken into consideration. Last but not least, the tight relationship between bonding behavior and crystal structure have not been well addressed. Consequently, it seems hard to acquire reliable electronic structures and related physical properties of plutonium borides. 

Several questions still remain to be answered. First of all, the mechanism of topological feature and its connection with strongly correlated 5$f$ electronic states. Secondly, the evolution pattern of itinerant to localized nature of 5$f$ electrons in plutonium borides. Thirdly, whether the orbital dependent correlations of 5$f$ electrons commonly exist in plutonium borides. To elucidiate the questions proposed above, it is of high priority to examine the fine electronic structures of plutonium borides to evaluate itinerant-localized dual nature, mixed-valence behavior, orbital selective 5$f$ electronic correlations and related crystal stability.

The remained parts of this paper are organized as follows. In Sec.~\ref{sec:method}, the computational details of DFT + DMFT approach are briefly presented. In Sec.~\ref{sec:results}, the quasiparticle bands, total and partial 5$f$ density of states, mixed-valence behavior and 5$f$ self-energy functions are discussed. The bonding behavior and crystal structure stability, as well as angular momentum coupling scheme are addressed. In the end, Sec.~\ref{sec:summary} gives a brief summary.


\section{Methods\label{sec:method}}
The DFT + DMFT method is established as a powerful approach to tackle with strongly correlated electronic systems, which has been widely applied to study the electronic structures of strongly correlated materials such as cerium-based heavy fermion materials and plutonium-based compounds. Essentially, the local quantum impurity model substitutes the local density approximation exchange-correlation effects in $f$ electrons, which integrates the advantage of realistic band structure calculation and many-body treatment of local interaction effects. To account for the strongly correlated 5$f$ electrons, we perform charge fully self-consistent calculations to explore the detailed electronic structure of PuB$x$ ($x$=1, 2, 6, 12) by employing the DFT + DMFT method which contains two main parts. The DFT part is solved by using the \texttt{WIEN2K} code~\cite{wien2k} which performs a full-potential linear augmented plane-wave (FP-LAPW) formalism. The DMFT part is solved by employing the \texttt{EDMFTF} package~\cite{PhysRevB.81.195107} which outputs high-precision total energy and force. 

In the DFT part, the experimental crystal structures for PuB$x$ ($x$=1, 2, 6, 12) were used throughout the calculations~\cite{Eick1965,McDonalda02824}. To study the paramagnetic states, we chose representative calculation temperature as $T \sim 116.0$~K (the inverse temperature $\beta = 100$). The generalized gradient approximation was adopted to formulate the exchange-correlation functional~\cite{PhysRevLett.77.3865}. Besides, the spin-orbit coupling was treated in a second-order variational manner. According to the geometry of crystal structure liste in table~\ref{tab:param}, the appropriate $k$-points' mesh was $15 \times 15 \times 15$ for PuB$x$ ($x$=1, 6, 12) and $17 \times 17 \times 12$ for PuB$_2$. In addition, $R_{\texttt{MT}}K_{\texttt{MAX}} = 8.0$ was utilized in the DFT calculation to meet the convergence criterion of charge and energy.
 
For the correlated 5$f$ electrons of plutonium, the DMFT part solves the quantum impurity model, which treats the plutonium atoms as a single impurity site and all 5$f$ orbitals are correlated. Above all, the electrostatic interaction was expressed as parametrical Slater's determinants~\cite{PhysRevB.59.9903}, so the Coulomb interaction was empirically chosen as $U = 5.0$~eV and the Hund's exchange $J_H=0.6$~eV which were diffusely used and suitable for plutonium-based compounds~\cite{PhysRevB.101.125123}. Furthermore, to subtract the double-counted term of interaction in the DFT part, we carefully selected the fully localized limit double-counting scheme to compute the impurity self-energy function~\cite{jpcm:1997}. In particular, we adopted the vertex-corrected one-crossing approximation (OCA) impurity solver~\cite{PhysRevB.64.115111} to solve the multi-orbital Anderson impurity models which greatly reduces the calculation burden compared to the hybridization expansion version of continuous-time quantum Monte Carlo quantum impurity solver (dubbed as CTQMC)~\cite{RevModPhys.83.349,PhysRevLett.97.076405} for comparison.  It should be pointed out that the size of operator matrix is $2^{14} \times 2^{14}$ for the 5$f$ electronic shell, which requires huge computer memory to save the data, and consumes considerable computing resources. To further reduce the computational resources required, the good quantum numbers $N$ (total occupancy) and $J$ (total angular momentum) were employed to classify the atomic eigenstates. Furthermore, the truncation was adopted by considering only those atomic eigenstates whose occupancy $N$ satisfying $N \in [N_{low}, N_{high}]$ will be taken into accounts in the local trace evaluation. Fortunately, a truncation ($N \in$ [3, 7]) is adequate for the local Hilbert space~\cite{PhysRevB.75.155113} to derive converged 5$f$ electronic states. 
To ensure the accuracy of the self-consistent calculation, the charge and energy convergence criteria were set as $10^{-5}$ e and $10^{-5}$ Ry, respectively. It is noteworthy that the advantage of OCA impurity solver lies in the real axis self-energy, the direct output $\Sigma (\omega)$ were used to obtain the momentum-resolved spectral functions $A(\mathbf{k},\omega)$ and density of states $A(\omega)$, as well as other physical observables.

\section{Results\label{sec:results}}

\begin{figure*}[th]
\centering
\includegraphics[width=\textwidth]{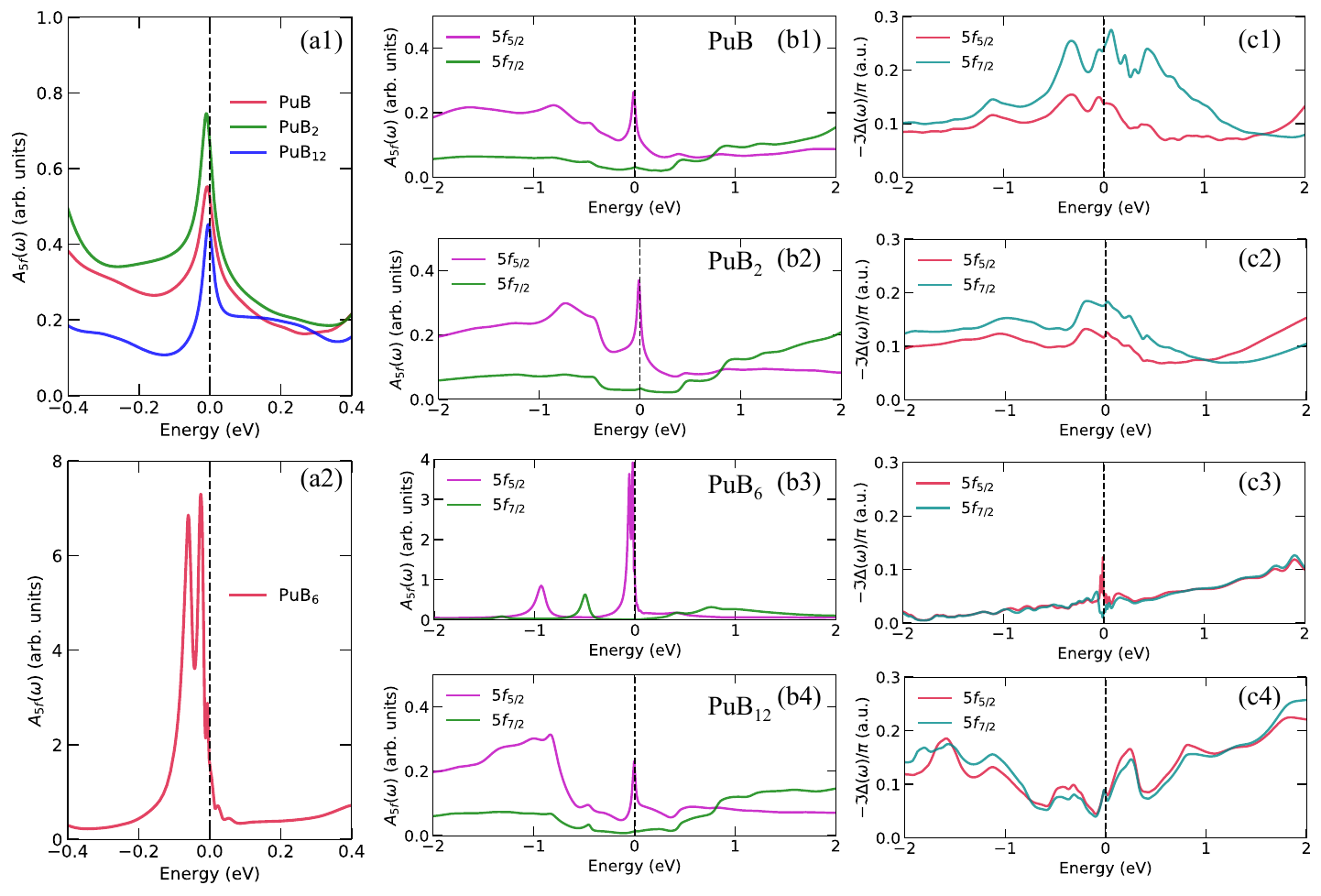}
\caption{(Color online). Electronic density of states and hybridization functions for PuB$x$ ($x$=1, 2, 6, 12) at 116 K obtained by DFT + DMFT method. (a1) Partial 5$f$ density of states for PuB, PuB$_2$, and PuB$_{12}$. (a2) Partial 5$f$ density of states for PuB$_6$. (b1)-(b4) The $j$-resolved 5$f$ partial density of states with $5f_{5/2}$ and $5f_{7/2}$ components represented by purple and green lines, respectively for PuB, PuB$_2$, PuB$_6$ and PuB$_{12}$, from top to bottom. (c1)-(c4) Hybridization functions denoted by red and cyan lines, respectively for PuB, PuB$_2$, PuB$_6$ and PuB$_{12}$, from top to bottom. 
\label{fig:dos}}
\end{figure*}

\subsection{Quasiparticle multiplets}

The pioneering advances have authenticated multiplet peaks near the Fermi level in Pu metal via observed photoemission spectroscopy and calculated quasiparticle multiplets, originating from 5$f$ valence fluctuations and atomic multiplets~\cite{PhysRevB.62.1773,PhysRevB.101.125123}. The atomic multiplets for 5$f$ electrons are believed to be generic in many Pu-based compounds~\cite{PhysRevLett.111.176404,PhysRevB.103.205134,Lu_2021}. In order to elucidate the quasiparticle multiplets and itinerant degree of 5$f$ electrons, it is imperative to investigate the evolution of 5$f$ correlated electronic states upon boron composition. Figure~\ref{fig:dos} delineates the density of states and hybridization functions of plutonium borides at 116 K. Figure~\ref{fig:dos}(a1) sketches the entire profile of 5$f$ density of states for PuB, PuB$_2$ and PuB$_{12}$, which share a similar single peak in the Fermi level but with slightly different spectral weights. Meanwhile, Fig.~\ref{fig:dos}(a2) shows 5$f$ density of states of PuB$_6$ which is distinctly different from the other three compounds and exhibits a much higher shoulder peak. A diminutive dip is discovered in the Fermi level which is associated with the onset of a pseudogap, in accordance with previous calculations~\cite{PhysRevLett.111.176404}. Additionally, two sharp peaks at --0.025 eV and --0.06 eV also emerge in the vicinity of Fermi level. 

The origin of these atomic multiplets is primarily encoded in the 5$f$ partial density of states [see Fig.~\ref{fig:dos}(b1)-(b4)]. Firstly, the 5$f$ states are split into 5$f_{5/2}$ and 5$f_{7/2}$ subbands stemming from spin-orbit coupling~\cite{RevModPhys.81.235,PhysRevB.102.245111,PhysRevB.101.125123}. A quasiparticle peak arises in the Fermi level, mostly belonging to 5$f_{5/2}$ orbital, which implicates metallic behavior for PuB, PuB$_2$ and PuB$_{12}$. Concurrently, two satellite peaks of PuB$_6$ at --0.9 eV and --0.5 eV with energy gap about 0.4 eV are attributed by 5$f_{5/2}$ and 5$f_{7/2}$ orbitals, respectively. In contrast, the reflected peaks above the Fermi level degrade into broad humps. Naturally, it is expected that the atomic multiplets are generated by 5$f$ valence state fluctuations, which leave signature features on the 5$f$ photoemission spectroscopy of PuB$_6$. Secondly, the quasiparticle weight of plutonium borides which are somewhat similar for PuB, PuB$_2$ and PuB$_{12}$ despite their diverse crystal structures. Thirdly, hybridization functions [see Fig.~\ref{fig:dos}(c1)-(c4)] depicts the hybridization strength between 5$f$ electrons and conduction bands ($c-f$). As can be seen, the strongest hybridization strength of PuB coincides with the shortest Pu-B distance 2.46 {\AA} in Tab.~\ref{tab:param}. In this scenario, the Pu-B distance regulated hybridization strength varies with boron composition.
 
\subsection{Electronic band structure}

\begin{figure*}[th]
\centering
\includegraphics[width=\textwidth]{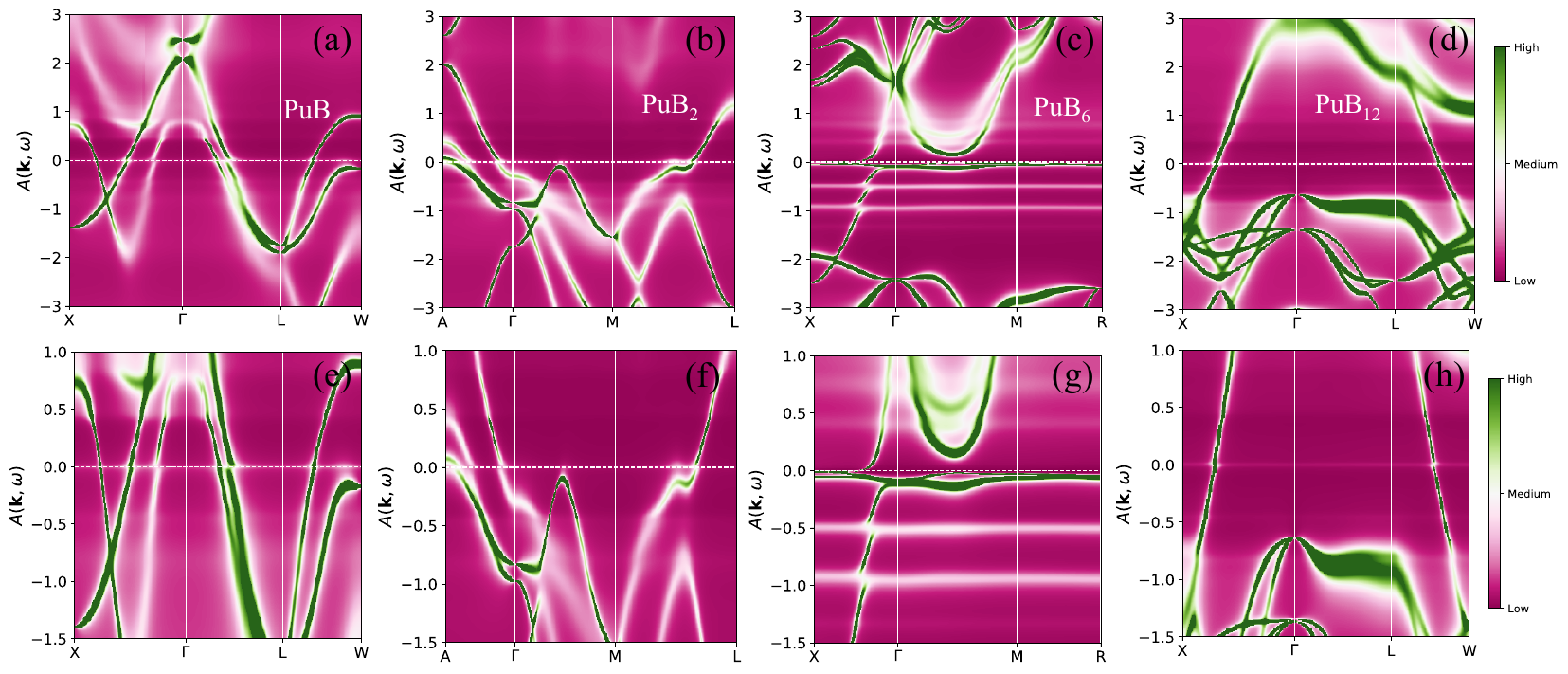}
\caption{(Color online). Momentum-resolved spectral functions $A(\mathbf{k},\omega)$ of plutonium borides at 116 K calculated by DFT + DMFT method. (a) PuB. The coordinates for the high-symmetry points are $X$ [0.0, 0.5, 0.5], $L$ [0.0, 0.5, 0.0], $W$ [0.5, 0.75, 0.25].
 (b) PuB$_2$. The coordinates for the high-symmetry points are $A$ [0.0, 0.0, 0.5], $M$ [0.5, 0.0, 0.0], $L$ [0.5, 0.5, 0.5].
 (c) PuB$_6$. The coordinates for the high-symmetry points are $X$ [0.5, 0.0, 0.0], $M$ [0.5, 0.5, 0.0], $R$ [0.5, 0.5, 0.5].
 (d) PuB$_{12}$. The coordinates for the high-symmetry points are the same with PuB.
An enlarged view of panel (e) PuB, (f) PuB$_2$, (g) PuB$_6$, (h) PuB$_{12}$ in the energy window $\omega \in [-1.5, 1.0]$ eV. The horizontal lines denote the Fermi level.
\label{fig:akw}}
\end{figure*}

\begin{figure*}[th]
\centering
\includegraphics[width=0.9\textwidth]{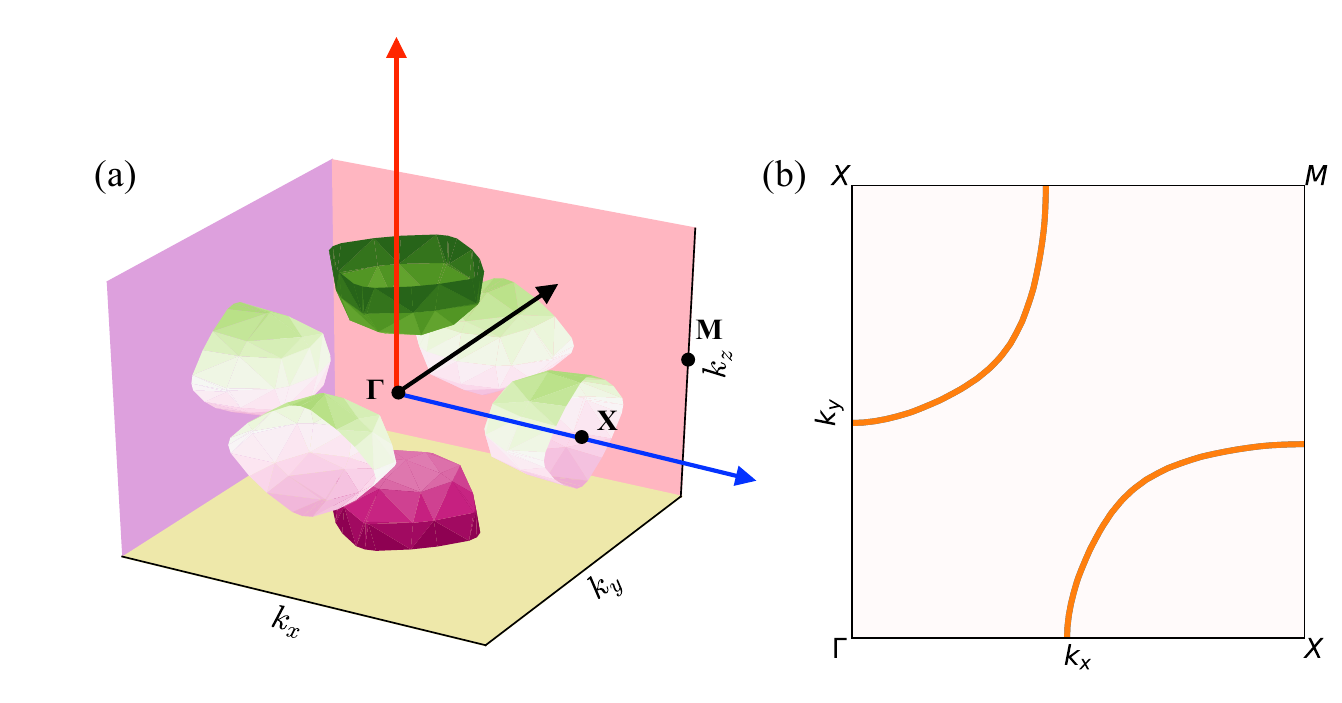}
\caption{(Color online). 
(a) Three-dimensional Fermi surface and (b) two-dimensional Fermi surface of PuB$_6$ at 116 K calculated by the DFT + DMFT method. Two-dimensional Fermi surface are on the $k_x-k_y$ plane (with $k_z = \pi/2$). 
\label{fig:FS}}
\end{figure*}

Now it is instructive to examine the momentum-resolved spectral functions which imprint intriguing features of plutonium borides [see Fig.~\ref{fig:akw}]. At first, the correctness of our computed electronic band structures is evaluated by referring to the band structure of PuB$_6$~\cite{PhysRevLett.111.176404} derived via adopting the continuous-time quantum Monte Carlo impurity solver. The overall band profile obtained from OCA and CTQMC impurity solvers share very semblable characteristics. As is shown in Fig.~\ref{fig:akw}(c), three salient flat narrow electronic bands parallel around the Fermi level which are contributed by 5$f$ states. Owing to the strong correlation among 5$f$ electrons, the electronic bands calculated by DFT + DMFT method are dramatically renormalized compared to the standard DFT results. After a close inspection, one may notice a weeny gap around the $X$ point just at the Fermi level [see Fig.~\ref{fig:akw}(g)], rooting from a band inversion between Pu-6$d$ and Pu-5$f$ orbitals. The band inversion hints a possible correlated topological insulator of PuB$_6$ which was affirmed by surface states analysis~\cite{PhysRevLett.111.176404} and motivates intriguing transport properties. Below the Fermi level, there exist two dispersionless 5$f$ electronic bands at --0.5 eV and --0.9 eV, respectively, which are split by spin-orbit coupling with an energy gap of 0.4 eV between 5$f_{5/2}$ and 5$f_{7/2}$ states. It should be pointed out that the energy gap of Pu-based compounds is larger than that of cerium-based compounds due to a heavier nucleus of Pu atom. Particularly, these narrow bands prominently hybridize with conduction bands along $X$ - $\Gamma$ high-symmetry line, opening evident hybridization gaps. Combined with the density of states in Fig.~\ref{fig:dos}, the conspicuous $c-f$ hybridization evinces the itinerant tendency of 5$f$ states. Moreover, the electronic struture could be visualized in the Fermi surface topology [see Fig.~\ref{fig:FS}(a)] which is usually probed by following dHvA quantum oscillation. Only one doubly degenerated bands cross the Fermi level (No. of bands: 24 and 25), which take six pretty parabolic shapes and agree with the band trait in Fig.~\ref{fig:akw}(c). It is noticable that a representative electron pocket appears along the $\Gamma$ - $M$ high-symmetry line in the energy region of 0 eV $\sim$ 1 eV which is attributed to conduction band. Despite the absence 5$f$ quasiparticle weight at --0.5 eV and --0.9 eV computed from CTQMC, the essential characteristics of energy bands are compliance with each other, corroborating the reliability of our results. A discrepancy in 5$f$ quasiparticle intensity below Fermi level might come from the overamplification of correlation effects in the OCA impurity solver.

Then we turn to the electronic band structures of PuB, PuB$_2$ and PuB$_{12}$. At first glance, 5$f$ electronic bands are not quite obvious as those in PuB$_6$ because of their small quasiparticle weight and faint band intensity. In the enlarged band structure near the Fermi level [see Fig.~\ref{fig:akw}(e), (f) and (h)], the conduction bands develop certain degree of bending when they meet 5$f$ states near the Fermi level. The advent of flat bands suggest the partially itinerant 5$f$ states. For PuB, these prominent conduction bands transect the Fermi level and mainly distribute in the energy range of --2 eV to 2 eV, inducing hole and electron pockets at $\Gamma$ and $L$ points, respectively.
For PuB$_2$, anomalous conduction bands mainly crowd below the Fermi level, revealing different nature from the other three compounds. Concerning PuB$_{12}$, two conduction bands with nearly linear dispersion locate about --1 eV to 3 eV, while most of the conduction bands gather below the Fermi level. It should be pointed out that the crystal structure of plutonium borides mediates the Pu-B bonding character, which in turn tunes the itinerant-localized nature of 5$f$ states. So far the calculated electronic band structure of plutonium borides serves as critical prediction for future angle-resolved photoemission spectroscopy (ARPES) experiment.

\subsection{Valence state fluctuations}

It is well established that $\delta$-Pu displays obvious mixed-valence behavior with noninteger occupation number deviating from nominal value 5.0. The 5$f$ electron atomic eigenstates obtained from the output of DMFT many-body states elaborate the valence state fluctuations and related mixed-valence behavior. Here $p_\Gamma$ is utilized to quantitatively describe the probability of 5$f$ electrons which stay in each atomic eigenstate $\Gamma$. Then the average 5$f$ valence electron is defined as $\langle n_{5f} \rangle = \sum_\Gamma p_\Gamma n_\Gamma$, where $n_\Gamma$ labels the number of electrons in each atomic eigenstate $\Gamma$. Finally, the probability of 5$f^n$ electronic configuration can be expressed as $\langle w(5f^{n}) \rangle = \sum_\Gamma p_\Gamma \delta (n-n_\Gamma)$. 

\begin{figure}[th]
\centering
\includegraphics[width=\columnwidth]{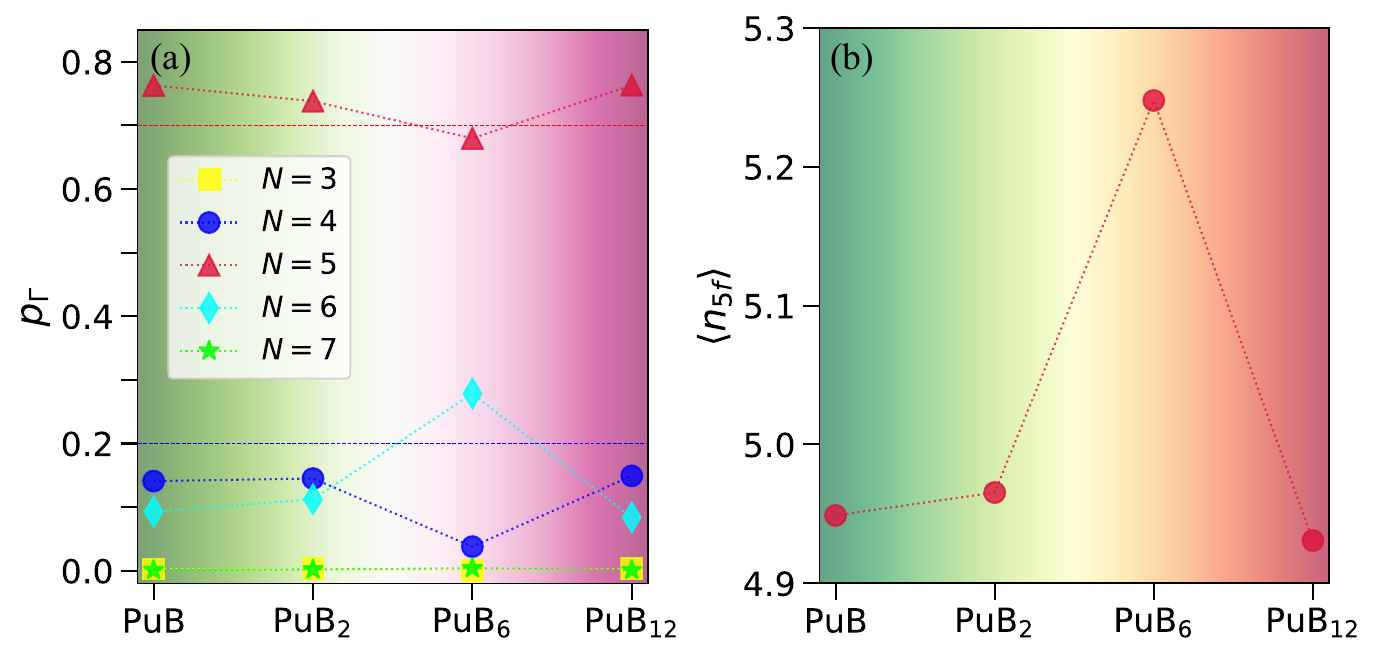}
\caption{(Color online). Valence state fluctuations in plutonium borides at 116 K computed by DFT + DMFT method. (a) Distribution probability of 5$f$ atomic eigenstates. (b) 5$f$ occupancy with respect to boron composition.
\label{fig:prob}}
\end{figure}

The calculated probabilities of 5$f^n$ electronic configuration for plutonium borides are visualized in Fig.~\ref{fig:prob}. Obviously, the probability of 5$f^5$ electronic configuration is prevailing, followed by a comparative proportion of 5$f^4$ and 5$f^6$ electronic configurations which fluctuate around 15\%. It is noticed that the contributions of 5$f^3$ and 5$f^7$ electronic configurations are too small to be seen similar to most plutonium compounds~\cite{PhysRevLett.111.176404,PhysRevB.103.205134}. 
According to the pattern of electronic configuration dependence on boron composition, plutonium borides are classified into two groups. PuB$_6$ stands out from the other three compounds with a considerably larger percentage of 5$f^6$ electronic configuration accounting for 27.8\%. Besides the predominant 5$f^5$ electronic configuration comes up to 67.9\%, the ratio of 5$f^4$ electronic configuration only limits to 3.8\%. The fact lies in that 5$f$ valence electrons tend to spend more time in the 5$f^5$ and 5$f^6$ electronic configurations rendering valence fluctuations and promoting quasiparticle multiplets. 
The regulated 5$f$ valence electron $\langle n_{5f} \rangle = 5.25$ is evocative of mixed-valence behavior in reminiscence of that in $\delta$-Pu~\cite{shim:2007}. On the other hand, the probability of 5$f^5$ electronic configuration approaches 80\%, followed by the comparable probabilities of 5$f^4$ and 5$f^6$ electronic configurations accounting for approximate 10\% of PuB, PuB$_2$, PuB$_{12}$. It means that a fairly smaller 5$f$ valence electron (less than 5.0) originates from a lower percentage of 5$f^6$ electronic configuration. Accordingly, the valence state fluctuations also distinguish PuB$_6$ from the other three compounds.

\subsection{Self-energy functions}

\begin{table}[th]
\caption{The electron effective mass $m^\star$ and quasi-particle weight $Z$ of $5f_{5/2}$ and $5f_{7/2}$ states for PuB$x$ ($x$=1, 2, 6, 12). \label{tab:sig}}
\begin{center}
\begin{tabular}{ccccc}
      & \multicolumn{2}{c}{$5f_{5/2}$} & \multicolumn{2}{c}{$5f_{7/2}$} \\
\hline
cases & $m^{\star}/m_e$ & $Z$ & $m^{\star}/m_e$ & $Z$ \\
\hline
PuB        & 51.762 & 0.019 & 6.869 & 0.146 \\
PuB$_2$    & 48.343 & 0.021 & 5.872 & 0.170 \\
PuB$_6$    & 11.587 & 0.086 & 18.047 & 0.055 \\
PuB$_{12}$ & 35.989 & 0.028 & 5.269 & 0.190 \\
\hline
\end{tabular}
\end{center}
\end{table}

Usually, the electronic correlations are encapsulated in the electron self-energy functions~\cite{RevModPhys.68.13,RevModPhys.78.865}. $Z$ denotes the quasiparticle weight or renormalization factor, which means the electronic correlation strength and can be acquired from the real part of self-energy functions via the following Eq.~\cite{RevModPhys.68.13}:
\begin{equation}
Z^{-1} = \frac{m^\star}{m_e} = 1 - \frac{\partial}{\partial \omega} \rm{Re} \Sigma(\omega) \Big|_{\omega = 0}. \label{eqsigma}
\end{equation}

The evaluated electron effective mass $m^\star$ and quasi-particle weight $Z$ for 5$f_{5/2}$ and 5$f_{7/2}$ states~\cite{RevModPhys.68.13} according to Eq.~\ref{eqsigma} are listed in table~\ref{tab:sig}. Provided that the self-energy functions are continuous and differentiable, the computed electron effective mass are accurate enough to derive three features. 
Firstly, the electron effective masses of PuB$_6$ with both 5$f_{5/2}$ and 5$f_{7/2}$ states are comparatively large, accompanied by diminutive renormalization factors, implying the moderate electronic correlations and intensive localization of 5$f$ states. The trend is in accordance with the low-energy electron scattering in the high-energy regime for PuB$_6$. Secondly, the electron effective masses of 5$f_{5/2}$ states are much larger than 5$f_{7/2}$ states for PuB, PuB$_2$ and PuB$_{12}$, which manifests heavily renormalized 5$f_{5/2}$ bands and strongly correlated 5$f_{5/2}$ states. Meanwhile, the strongly correlated 5$f_{5/2}$ states exhibit obvious peaks in the Fermi level, manifesting the metallic behavior. On the contrary, less renormalized 5$f_{7/2}$ states suggest relatively weakly correlated 5$f_{7/2}$ states. Thirdly, slight variations of electron effective masses among PuB, PuB$_2$ and PuB$_{12}$ result from diverse band renormalization strengths. Therefore electronic correlation strength and itinerant degree of freedom are distinct for 5$f_{5/2}$ and 5$f_{7/2}$ states, which are related with the substantive correlated electronic characteristic, further verifying the orbital dependent correlated states.

\section{Discussions\label{sec:dis}}
In this section, both itinerant-localized dual nature and strongly correlated 5$f$ electronic state are explored of plutonium borides to unravel the bonding behavior and underlying mechanism of crystal stability. 

\textbf{Bonding behavior and crystal stability.}
It is generally believed that itinerant 5$f$ states prefer to hybridize with conduction bands and open hybridization gaps. Subsequently, noteworthy quasiparticle multiplets and significant valence state fluctuations along with strong mixed-valence behavior are often visualized in such system. Just as the case for temperature driven itinerant 5$f$ electron is commonly weakly correlated, you may wonder whether the situation holds in pressure driven or chemical doping 5$f$ electron systems. Take PuB$_6$ for example, the largest Pu-B distance of the four compounds [see Tab.~\ref{tab:sig}] stretches the Pu-B bonding radius so as to weaken the wavefunction overlap between Pu and B atoms.
Hence the attenuated bonding strength may boost localized 5$f$ electrons, which interprets a pseudogap at the Fermi level in the calculated density of states. It is emphasized that the potential topological feature is tightly related to the correlated 5$f$ orbitals which hybridize with 6$d$ oribtals to induce the band inversion, corroborating a correlated topological insulator of PuB$_6$. 

As is mentioned above, partial itinerancy of 5$f$ states embody the $c-f$ hyridization gap below the Fermi level, finite quasiparticle weight, evident quasiparticle multiplets and remarkable valence state fluctuations. In this respect, 5$f$ states reveal itinerant-localized dual nature which serve as a long-lived issue in condensed matter physics. On the other hand, electron effective masses derived from self-energy functions implicate medium electronic correlation of both $J=5/2$ and $J=7/2$ states, which is in conformity with the pattern in temperature driven itinerant 5$f$ states.
In contrast to PuB$_6$, the observed band intensity of PuB, PuB$_2$ and PuB$_{12}$ is not quite prominent near the Fermi level together with moderate valence state fluctuations, hinting pretty strongly correlated 5$f$ electronic states. Actually, an established picture portrays that itinerant 5$f$ states are moderately correlated, while localized 5$f$ states are inclined to be strongly correlated in most temperature and chemical doping 5$f$ electron systems.

Even so, some delicate differences still exist in quasiparticle weight [see Fig.~\ref{fig:dos}(b1)-(b4)] and hybridization strength [see Fig.~\ref{fig:dos}(c1)-(c4)]. It is noted that the spectral weight obeys the sequence as PuB$_2$ $\textgreater$ PuB $\textgreater$ PuB$_{12}$, which accords with the series derived from hybridization strength. It is illuminating to examine the detail of crystal structure parameters listed in table~\ref{tab:param}. Interestingly, the typical Pu-Pu distance follows the order as PuB$_2$ $\textless$ PuB $\textless$ PuB$_{12}$, which stresses the atomic distance affecting the itinerant degree of freedom for 5$f$ electrons. Furthermore, 5$f$ electron occupancy signifies the mixed-valence nature which could be measured by the electron energy-loss spectroscopy and X-ray absorption spectroscopy~\cite{Booth26062012}. Actually, when 5$f$ and 6$d$ orbitals of Pu atom hybridize with the 2$s$ and 2$p$ orbitals of boron atom, the fluctuating 5$f$ electron enables Pu-B bonding, which subsequently alters the charge distribution and finally influences the crystal structure stability of plutonium borides. The hypothesis could be generalized to understand the stability of Th- and Am-diborides~\cite{Eick1969}. The lack of 5$f$ electron for Th atom makes it hard to bond with 2$p$ orbital of boron atom so that the poor-boron compound like ThB$_2$ hardly exists in nature. Likewise, a large element number increases the atomic distance, which induces a more localized 5$f$ state. The weak bonding between Am and B atom is unable to stabilize AmB$_2$, where boron composition is rather low in diboride.

\textbf{Angular momentum coupling scheme.}
The evolution pattern of 5$f$ electrons occupancy in the $5f_{5/2}$ and $5f_{7/2}$ levels across actinide series is a crucial question. Commonly, it is determined by the angular momentum coupling scheme of each actinide. For the multielectronic systems, there exist three ways of angular momentum coupling on account of the relative strength of spin-orbit coupling and electronic interaction. Namely, Russell-Saunders (LS) coupling, $jj$ coupling, and intermediate coupling (IC)~\cite{RevModPhys.81.235}. For the ground states of late actinides, intermediate coupling scheme is usually the favorite. A natural question rises with the 5$f$ orbital occupancy and angular momentum coupling scheme for PuB$x$ ($x$=1, 2, 6, 12). 

Since X-ray absorption spectroscopy is a powerful technique to detect the electronic transitions between core 4$d$ and valence 5$f$ states, it is used to observe the occupancy of 5$f$ electrons of actinides.
 The strong spin-orbit coupling for the 4$d$ states leads to two absorption lines, representing the $4d_{5/2} \rightarrow 5f$ and $4d_{3/2} \rightarrow 5f$ transitions, respectively. The X-ray absorption branching ratio $\mathcal{B}$ is defined as the relative strength of the $4d_{5/2}$ absorption line~\cite{PhysRevA.38.1943}. It calibrates the spin-orbit coupling interaction strength in 5$f$ shell. Under the approximation that the electrostatic interaction between core and valence electrons is neglected, the expression for $\mathcal{B}$ is written as~\cite{shim:2007}:
\begin{equation}
\label{eq:ratio}
\mathcal{B} = \frac{3}{5} - \frac{4}{15} \frac{1}{14 - n_{5/2} - n_{7/2}} \left ( \frac{3}{2} n_{7/2} - 2 n_{5/2} \right ),
\end{equation}
where $n_{7/2}$ and $n_{5/2}$ are the 5$f$ occupation numbers for the $5f_{7/2}$ and $5f_{5/2}$ states, respectively.
The calculated results are listed in table~\ref{tab:n5f}, giving $\mathcal{B}$(PuB$_6$) $\textgreater$ $\mathcal{B}$(PuB$_{12}$) $\textgreater$ $\mathcal{B}$(PuB) $\approx$ $\mathcal{B}$(PuB$_2$). That means the angular momentum coupling scheme of four compounds are assigned to intermediate coupling, which is similar to that in heavy actinides Pu~\cite{PhysRevB.101.125123} and Cm~\cite{PhysRevB.101.195123}.

\begin{table}[th]
\caption{The weights for 5$f$ electronic configurations ${w}(5f^n)$, 5$f$ orbital occupancy ($n_{5/2}$, $n_{7/2}$, and $n_{5f}$), and X-ray absorption branching ratio $\mathcal{B}$ for PuB$x$ ($x$=1, 2, 6, 12). \label{tab:n5f}}
\begin{center}
\begin{tabular}{cccccccc}
\hline
cases & ${w}(5f^4)$ & ${w}(5f^5)$ & ${w}(5f^6)$ & $n_{5/2}$ & $n_{7/2}$ & $n_{5f}$ & $\mathcal{B}$ \\
\hline
PuB        & 14.09\% & 76.28\% & 9.27\% & 3.77 & 1.17 & 4.94 & 0.7706 \\
PuB$_2$    & 14.49\% & 73.80\% & 11.22\% & 3.77 & 1.19 & 4.96 & 0.7702 \\
PuB$_6$    & 3.79\% & 67.96\% & 27.84\% & 4.02 & 1.23 & 5.25 & 0.7888 \\
PuB$_{12}$ & 14.91\% & 76.30\% & 8.38\% & 3.78 & 1.15 & 4.92 & 0.7714 \\
\hline
\end{tabular}
\end{center}
\end{table}

\section{Conclusions\label{sec:summary}}
The detailed electronic structures of plutonium borides embracing momentum-resolved spectral functions, density of states, valence state fluctuations and self-energy functions are comprehensively studied by employing the traditional density functional theory combined with single-site dynamical mean-field approach. It is found that PuB$_6$ is distinguished from PuB$_x$ ($x$=1, 2, 12), evincing topological feature with mixed-valence behavior. Meanwhile, the metallic PuB$_x$ ($x$=1, 2, 12) implicate tunable itinerancy of 5$f$ electrons which is virtually affected by $c-f$ hybridization. By tuning boron composition, the hybridization strength varies against Pu-B distance, which might be the vital factor in the growing localization of 5$f$ states along with enlarging Pu-B distance. The regulated charge distribution and bonding behavior contribute to the lattice stability of plutonium borides. Actually, itinerant-localized nature interconnecting with 5$f$ electronic correlation means that localized 5$f$ states is probably strongly correlated. Overall, the calculated electronic band structure and density of states serve as crucial predictions which deserve further experimental corroboration.

\section*{Acknowledgments}
This paper was supported by the National Natural Science Foundation of China (Grants No.~11874329, No.~11934020), CAEP Project (No.~TCGH0710). 
\section*{References}

\bibliographystyle{unsrt}
\bibliography{PuBx}

\end{document}